\documentclass{iopart}
\usepackage{graphicx}

\begin{document}
\title[Phase transition in CoS$_2$ at high pressures]{Thermodynamics of the ferromagnetic phase transition in nearly half metallic CoS$_2$ at high pressures}

\author{F S Elkin$^1$, I P Zibrov$^1$, A P Novikov$^1$, S S Khasanov$^2$, V A Sidorov$^1$, A E Petrova$^1$, T A Lograsso$^3$, J D Thompson$^4$, S.M. Stishov$^1$}
\address{$^1$Institute for High Pressure Physics of Russian
Academy of Sciences, Troitsk, Russia}
\address{$^2$Institute for Solid State Physics, Russian Academy of Sciences, Chernogolovka, Moscow Region, Russia}
\address{$^3$Ames Laboratory, Iowa State University, Ames, IA 50011, USA}
\address{$^4$Los Alamos National Laboratory, Los Alamos, NM 87545, USA}
\ead{sergei@hppi.troitsk.ru}
\begin{abstract}
The volume change and heat capacity at the ferromagnetic phase transition in CoS$_2$ were measured at high pressures using X-rays generated by the Argonne synchrotron light source and by ac-calorimetry, respectively. The transition entropy, calculated on the basis of these experimental data, drops along the transition line due to quantum degradation, as required by Nernst's law. The volume change increases strongly along the transition line, which is explained by specifics of the compressibility difference of coexisting phases that results from nearly half metallic nature of the ferromagnetic phase of CoS$_2$.
\end{abstract}

\section{Introduction}
Cobalt disulphide, CoS$_2$, a metallic compound with the cubic pyrite-type crystal structure~\cite{1}, exhibits a phase transition to a ferromagnetic state at $T_c\sim122$ K Ref.~\cite{2} that magnetic and electric properties indicate to be itinerant in nature~\cite{2,3,4,5,6}. Upon entering the ferromagnetic state CoS$_2$ becomes a nearly half metal with a significant decrease in the density of states at the Fermi level that is reflected in an increase in the resistivity below $T_c$~\cite{7,8}. The temperature-pressure phase diagram of CoS$_2$ has been studied to high pressures where experiments show that the ferromagnetic transition decreases monotonically and trends to zero at $\sim$6 GPa~\cite{9,10,11,12}. A change from continuous to first-order phase transition with a tricritical point close to ambient pressure was suggested in Refs.~\cite{11,12}, which accounts for the absence of a non-Fermi liquid temperature dependence of the resistivity near this critical pressure~\cite{12}.  The conclusion for a strong first-order quantum-phase transition in CoS$_2$ was based on indirect arguments~\cite{12}; consequently,  direct thermodynamic studies are needed to shed light on the evolution and properties of the phase transition in CoS$_2$ at high pressures.

\section{Experimental}
\begin{figure}[htb]
\includegraphics[width=80mm]{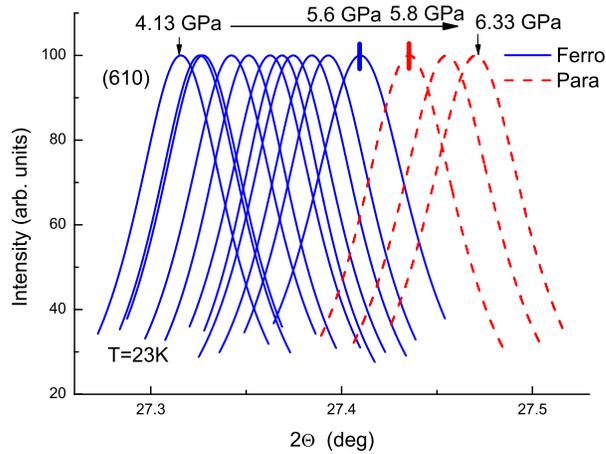}
\caption{\label{fig1} (Color online) Position of (601) reflection from a single crystal of  CoS$_2$ as a function of pressure at 23 K. Reflection shift between 5.6 and 5.8 GPa corresponds to the first order phase transition.}
 \end{figure}

We report an X-ray study of the lattice parameter change at the phase transition in CoS$_2$ along the pressure-dependent transition line. Single crystals of CoS$_2$ were grown by chemical-vapor-transport~\cite{12}, and some crystals were ground in an agate mortar to prepare a powder sample. X-ray diffraction studies of both crystals and powders of CoS$_2$ were performed at the HPCAT (16BM-D) beam line of the Advanced Photon Source (APS), Argonne National Laboratory. Single-crystal data were collected using the rotation method ($\omega$-axis rotation rate of $\pm17^\circ/500$ sec, X-ray wave length $\lambda=0.424603 {\AA}$) and unit-cell parameters were calculated from the (610) and (440) reflections. In the powder-diffraction experiments, the X-ray wave length $\lambda=0.354300 {\AA}$ was chosen to get the ten strongest reflections. In both experiments, an image plate detector MAR345, calibrated with fine CeO$_2$ standard powder, was used for data collection. Examples of the diffraction patterns are given in Fig.~\ref{fig1}. The collected data were subjected to the full profile analysis using the GSAS software package~\cite{Larson,Toby}, with a resulting accuracy of the unit-cell parameter determination of $\pm1\times10^{-4} {\AA}$.

For the diffraction experiments, high pressures were generated in a diamond anvil cell with $\sim$400 $\mu m$ culet diameters and $\sim60^\circ$ aperture. A 200 $\mu m$ hole was drilled in the pre-indented stainless-steel gasket. A powder or single crystal sample of CoS$_2$ ($\sim50\times50\times10\ \mu m^3$ in size) and ruby chips were placed in the sample chamber that was filled with helium to a pressure of $\sim$200 MPa. A gas-membrane device equipped with a pressure-control system was employed to pressure-load the cell. For these experiments, the cell was attached to a cold-finger type cryostat, which provided temperatures down to 15 K measured by a Cernox thermometer. Pressure was measured by the ruby luminescence technique with accuracy $\pm2\times10^{-5}$ GPa making use of the standard ruby calibration scale and with the appropriate temperature correction, in correspondence with procedures accepted in the HPCAT.

The X-ray study was supplemented by ac specific heat measurements at high pressures~\cite{14Sidorov}, which were generated in a miniature clamped toroid-type anvil pressure cell~\cite{15Petrova} with a glycerol/water (3:2 by volume) mixture as the pressure medium. Pressure at low temperatures was estimated from the superconducting transition temperature of lead~\cite{16Eiling}.

\section{Results}
\begin{figure}[htb]
\includegraphics[width=80mm]{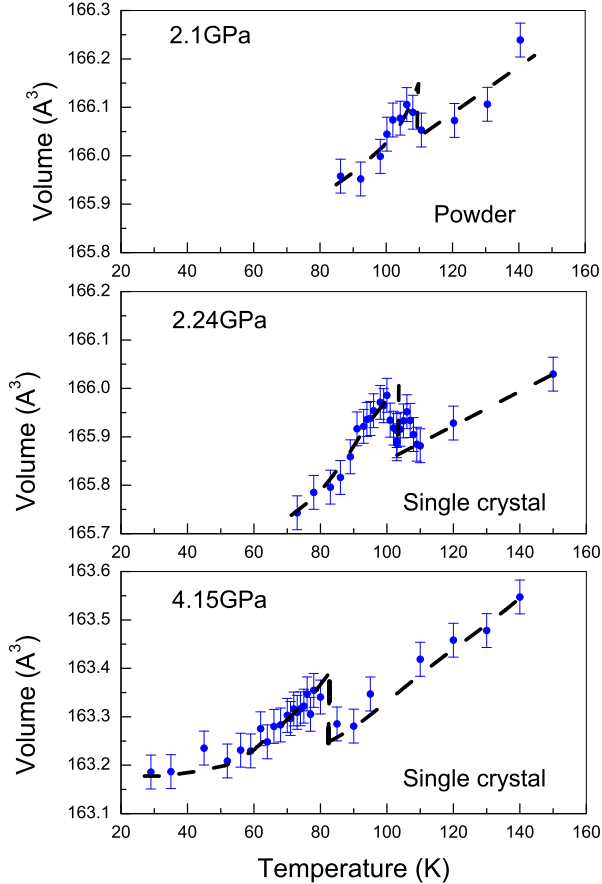}
\caption{\label{fig2} (Color online) Examples of the temperature dependence of the unit-cell volume of CoS$_2$ in the vicinity of its ferromagnetic phase transition. Dashed curves are guides to the eye. It is tempting to ascribe the seemingly two anomalies at the phase transition shown in the middle panel to a splitting of the phase transition; however, heat capacity data do not support this supposition.}
 \end{figure}

\begin{figure}[htb]
\includegraphics[width=80mm]{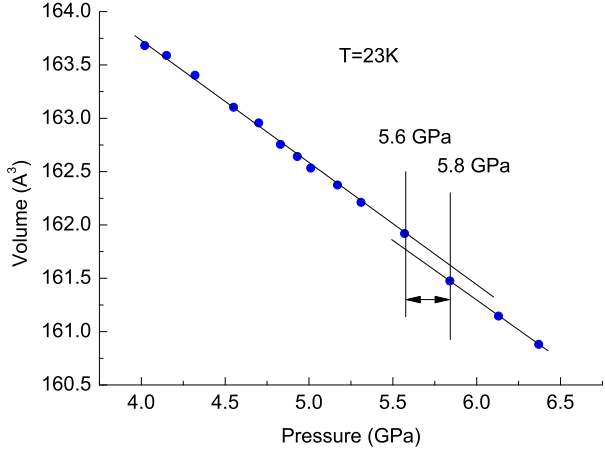}
\caption{\label{fig3} (Color online) Compression isotherm of CoS$_2$. As shown by the discontinuity in these data, a first order phase transition occurs between 5.6 and 5.8 GPa. Error bars correspond to the circle size.}
 \end{figure}

\begin{figure}[htb]
\includegraphics[width=80mm]{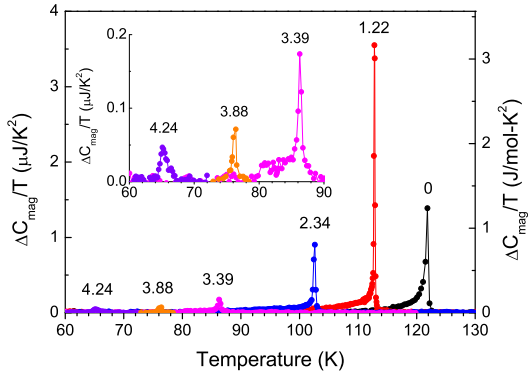}
\caption{\label{fig4} (Color online) Heat capacity of CoS$_2$ in the vicinity of its phase transition. Numbers above the peaks correspond to pressure values in GPa.
In the inset the heat capacity peaks at 3.39, 3.88 and 4.24 GPa are shown in the enlarged scale}
 \end{figure}

\begin{figure}[htb]
\includegraphics[width=80mm]{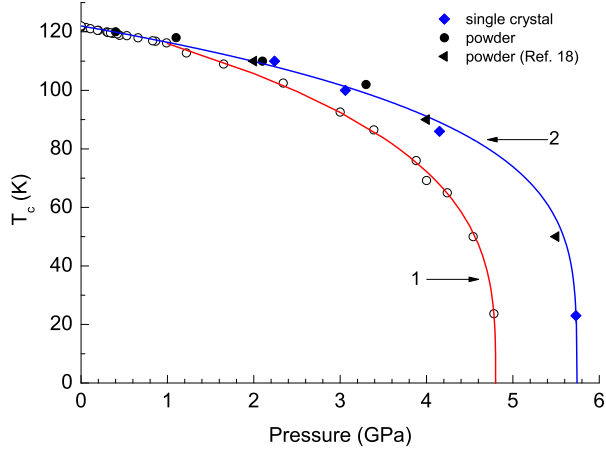}
\caption{\label{fig5} (Color online) Phase diagram of CoS$_2$ at high pressure. Curve 1 was determined making use a "lead" manometer to measure pressure~\cite{12}. Curve 2 was determined by X-ray diffraction and the pressure measured by a ruby sensor.}
 \end{figure}

\begin{figure}[htb]
\includegraphics[width=80mm]{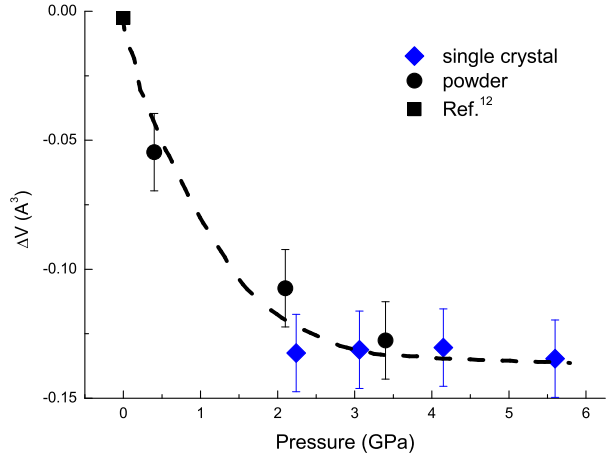}
\caption{\label{fig6} (Color online) Volume change at the ferromagnetic phase transition in single crystal and powder samples of CoS$_2$ as a function of pressure. The dashed curve is a guide to the eyes.}
 \end{figure}

\begin{figure}[htb]
\includegraphics[width=80mm]{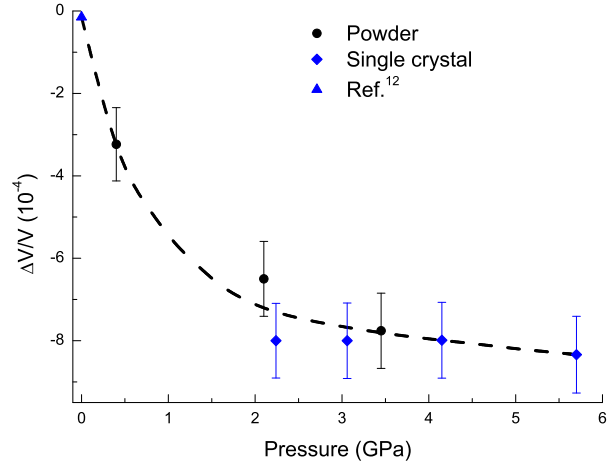}
\caption{\label{fig7} (Color online) Relative volume change at the ferromagnetic phase transition in CoS$_2$ as a function of pressure. The dashed curve is a guide to the eyes.}
 \end{figure}

Figure~\ref{fig2} gives typical results for the variation in unit-cell volume around the phase boundary. Experimental conditions, unfortunately, introduce sufficient error in accurately determining the pressure and temperature that it is not possible to distinguish in these data the difference between a continuous phase transition and a jump inherent to a first-order phase transition.  We take the change in cell volume at the phase transition to be given by the dashed curves in Fig.~\ref{fig2}. Consequently, it is not possible to draw conclusions from these data about the tricritical behavior in CoS$_2$ (Ref.~\cite{11,12}). On the other hand, these experimental uncertainties are inconsequential in determining the isothermal change in cell volume as a function of pressure in the high pressure regime that is plotted in Fig.~\ref{fig3}.  In the next section, results of Figs.~\ref{fig2} and ~\ref{fig3} will be compared to heat capacity data plotted in Fig.~\ref{fig4}. These data demonstrate an evolution of the phase-transition heat with temperature and pressure. The initial rise of the heat capacity peak with increasing pressure probably signifies the crossover from second to first order transitions.
The pressure dependence of the phase-transition temperature deduced from the current experimental data is shown in Fig.~\ref{fig5}. As seen, the phase line obtained from X-ray diffraction differs significantly from that determined from electrical resistivity, susceptibility and heat capacity data ~\cite{12}.  Obviously, calibrations of the low temperature ruby pressure scale and the scale based on the superconducting transition temperature of lead greatly disagree. It is worth noting that the phase-transition line in CoS$_2$ that also had been established in Ref.~\cite{10,12} with use the "lead" high pressure scale also differs from the current data probably due to non hydrostatic experimental conditions. Figures~\ref{fig6} and ~\ref{fig7} summarize the pressure-dependent evolution of $\Delta V$ and $\Delta V/V_F$ along the transition line. Here $\Delta V =V_P-V_F$, where $V_P$ and $V_F$ are the unit-cell volume calculated from lattice-parameter data in the paramagnetic and ferromagnetic phases, respectively.  We emphasize a special behavior of $\Delta V$: its absolute value  $|\Delta V|$ increases with pressure, which probably indicates involvement of some nontrivial physics. Despite an increase of $\Delta V$ by an order of magnitude in the range from 120 to 20K, the maximum ratio $\Delta V/V_F$ reaches only $\sim0.1$ percent, so there is not a strong first order phase transition in CoS$_2$ at high pressure.


\begin{figure}[htb]
\includegraphics[width=80mm]{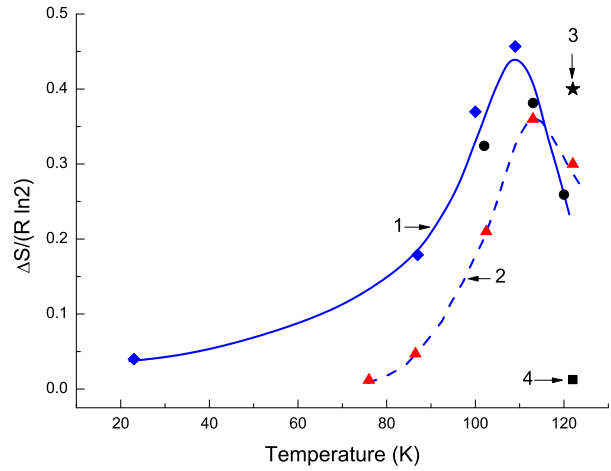}
\caption{\label{fig8} (Color online) Entropy of the phase transition in CoS$_2$ as a function of temperature. Symbols denote the entropy measured at the phase transition and dashed curves are guides. Curve1: entropy change calculated from the Clapeyron-Clausius equation; curve 2:  entropy change calculated from heat capacity data (Fig.~\ref{fig4}); point 3: calorimetry result at ambient pressure~\cite{13Ogawa}; point 4: calculated from dilatometric measurements at ambient pressure~\cite{12}. Pressure measurements in experiments leading to curves 1 and 2 were based on different pressure scales (Fig.~\ref{fig5}). To compare these results on an equal footing, they were plotted as functions of temperature.}
 \end{figure}
 \begin{figure}[htb]
\includegraphics[width=80mm]{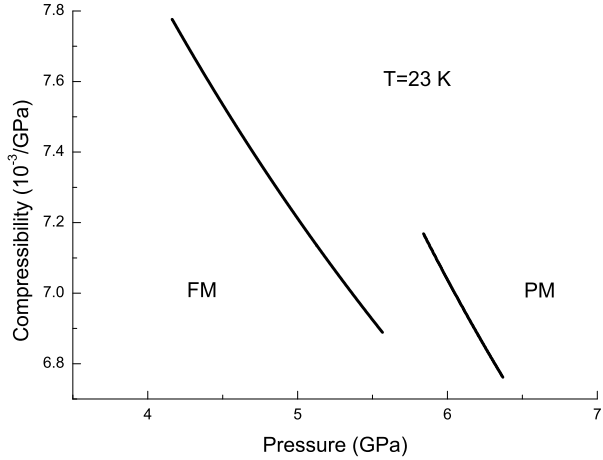}
\caption{\label{fig9}  Compressibility discontinuity at the phase transition in CoS$_2$ at high pressure, calculated from the compression isotherm (Fig.~\ref{fig3}). The compressibility of paramagnetic phase exceeds that of the ferromagnetic phase.}
 \end{figure}
 \begin{figure}[htb]
\includegraphics[width=80mm]{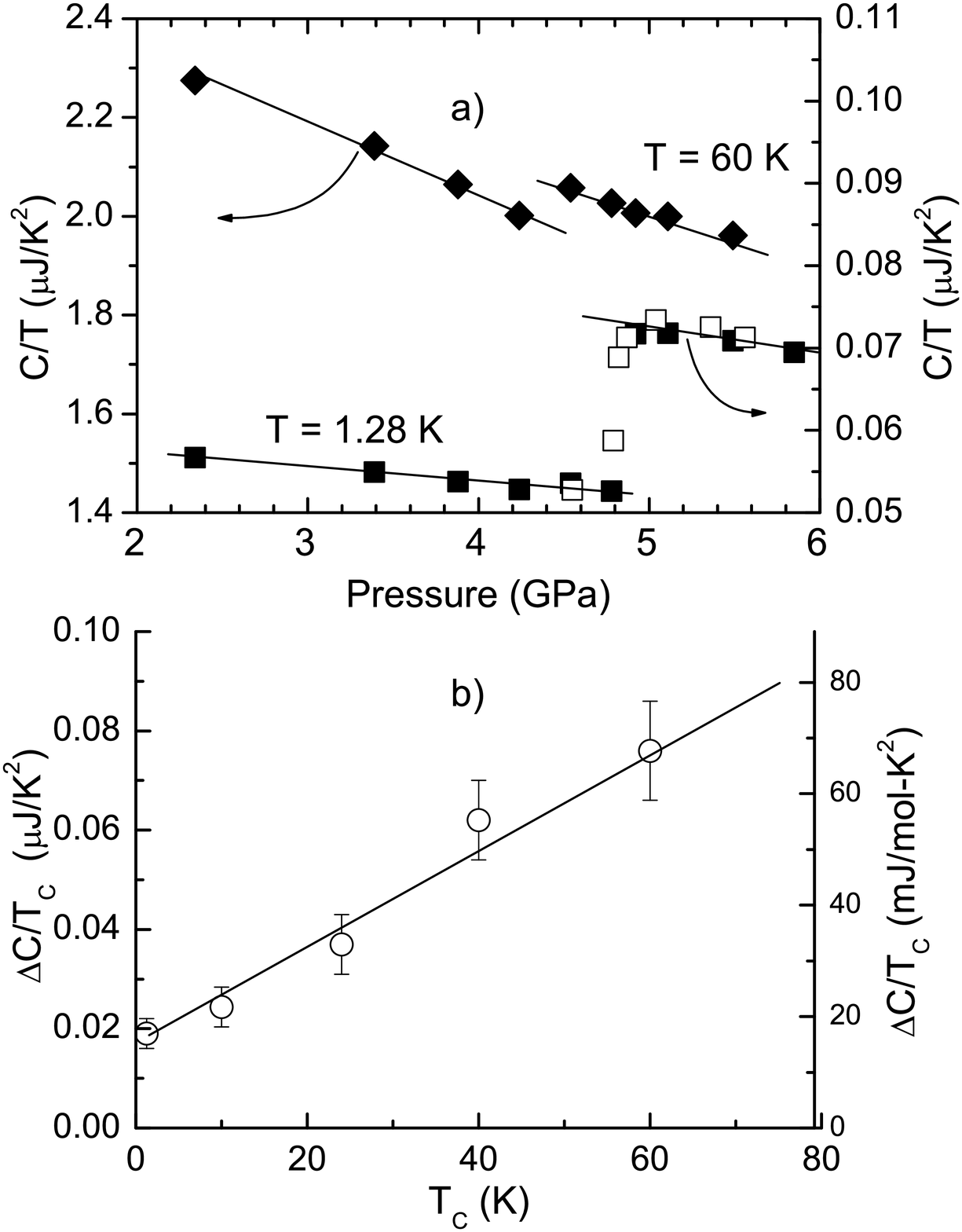}
\caption{\label{fig10}  a) Jumps of heat capacity divided by temperature at the phase transition in CoS$_2$ at T=1.28 K and T= 60 K. Values of $C_p/T$ correspond to heat capacity of the sample plus uncertain amount of pressure transmitting media. Heat capacity of the media stays constant at the phase transition therefore providing a possibility to calculate an absolute value of $\Delta C_p/T$, as shown in the figure. Note that the jump at 60 K is much higher than at 1.28 K  b) Jumps of heat capacity divided by temperature at the phase transition in CoS$_2$ as a function of transition temperature $T_c$. Heat capacity of the media stays constant at the phase transition therefore providing a possibility to calculate an absolute value of $\Delta C_p/T$, as shown in the figure. }
 \end{figure}

The presented data permit calculations of the entropy of the phase transition in CoS$_2$.  Two sets of calculations are presented in Fig.~\ref{fig8}. Set 1 is calculated from the volume change at the phase transition (Fig.~\ref{fig6}) together with the Clausius-Clapeyron equation $dT/dP = \Delta V/\Delta S$, which connects the slope of the transition curve with a difference of volume and entropy. Set 2 was obtained by integrating the heat capacity data (Fig.~\ref{fig4}). Both curves in Fig.~\ref{fig8} qualitatively agree, reflecting a fast quantum degradation of the transition entropy with lowering temperature from temperatures as high as 110 K. The quantitative difference between the two sets is attributed to many factors, including uncertainty in the heat capacity of CoS$_2$ due to inadequate subtraction of a contribution from the pressure transmitting medium, accuracy in determination the volume change, the slope of the phase transition line, etc..

\section{Discussion}
As seen from the experimental data, there is a steep growth of the volume change with pressure  at the ferromagnetic phase transition in CoS$_2$. This behavior could be connected at least partly with the tricritical crossover supposedly observed in CoS$_2$  at its magnetic phase transition close to ambient pressure~\cite{11,12}. In this case, first order features should grow on compression, which is in fair agreement with the general evolution of $\Delta V$. On the other hand, a negative volume change at a phase transition, such as seen in CoS$_2$, is typical of itinerant magnets and is a manifestation of a magnetovolume effect or volume magnetostriction~\cite{18}. Neglecting the spin fluctuations, the volume change due to band polarization and associated spontaneous magnetization  can be written as~\cite{19,20}
$$\Delta V/V_F=(C/K)M^2,$$
where M is the magnetization, K is the bulk modulus, and C is the magnetovolume coupling constant. As seen in Fig.~\ref{fig7}, $\Delta V/V_F$ changes by an order of magnitude along the transition line. To explain this observation, the magnetization M of the ferromagnetic phase CoS$_2$ at the transition line should increase considerably with pressure, which seems implausible and contradicts the experimental data and calculations~\cite{21,22}. At the same time, from $\Delta V$ along the phase-transition line it necessarily follows that the compressibility of the ferromagnetic phase is always lower than that of the paramagnetic phase. This conclusion is clearly supported by the compressibility calculated from the compression isotherm (Fig.~\ref{fig9}). This situation seems to be not common. For instance, in the case of the helical magnet MnSi a reverse relationship is observed between compressibilities of the coexisting magnetic and paramagnetic phases~\cite{23}. But, the nearly half metal nature of the ferromagnetic phase of CoS$_2$ creates a new situation. The point is that most of the electrons of the minority band become localized in the nearly half metallic state. This influences the repulsive interaction between electrons in CoS$_2$ in such a way to decrease the compressibility. An analogous situation is realized in a high pressure study of the model half-metal CrO$_2$~\cite{24}. This conclusion is supported by the pressure-dependent heat capacity at 1.28 K, which shows (Fig.~\ref{fig10}) a positive jump at the transition from ferromagnetic to paramagnetic states. At this low temperature, phonon and magnon degrees of freedom are practically frozen out. As a result, electronic contributions dominate $C/T$ and therefore the jump indicates an increase of the electronic density of states in the paramagnetic phase, as expected for a transition from half metallic to true metallic states. Note that at higher temperatures the corresponding jumps increase significantly ((Fig.~\ref{fig10}b) that may imply decreasing half metallicity with pressure if the phonon and magnon contributions can be neglected.

It should be mentioned that in the Landau theory compressibility and heat capacity changes at a phase transition have the same sign, which seemingly contradicts the current observation~\cite{25}. In the Landau theory, however, only anomalous parts of thermodynamic quantities are considered; whereas, the regular background contributions, which could drastically change at a phase transition, cannot be treated in a general way. In CoS$_2$, the sign of the compressibility change is certainly defined by the background, which eliminates a contradiction with the Landau theory.
\section{Conclusion}
The volume change and heat capacity were measured at the ferromagnetic phase transition in CoS$_2$ at high pressure, and the transition entropy was calculated from these experimental data. The transition entropy drops along the transition line due to quantum degradation, as required by the Nernst law. The volume change increases substantially along the transition line and is explained by specifics of the compressibility difference of the coexisting phases, which results from the nearly half metallic nature of the ferromagnetic phase of CoS$_2$.

\ack
This work was supported by the Russian Foundation for Basic Research (grant 12-02-00376-a), Program of the Physics Department of RAS on Strongly Correlated Electron Systems and Program of the Presidium of RAS on Strongly Compressed Matter. Work at Los Alamos National Laboratory was performed under the auspices of the U.S. Department of Energy, Office of Basic Energy Sciences, Division of Materials Sciences and Engineering. T.A.L. wish to acknowledge research performed at Ames Laboratory. Ames Laboratory is operated for the U.S. Department of Energy by Iowa State University.
A portion of this work was performed at HPCAT (Sector 16), Advanced Photon Source (APS), Argonne National Laboratory. HPCAT operations are supported by DOE-NNSA under Award No. DE-NA0001974 and DOE-BES under Award No. DE-FG02-99ER45775, with partial instrumentation funding by NSF. APS is supported by DOE-BES, under Contract No. DE-AC02-06CH11357. F.E and I.Z  greatly appreciate help of C. Kenney-Benson, D. Ikuta and D.Popov.


\section*{References}
\begin{thebibliography}{99}
\bibitem{1}	Elliot N  1960 {\it J. Chem. Phys.} \textbf{33} 903
\bibitem{2} Jarrett H S, Cloud W H, Bouchard R J, Butler S R, Frederick C G, Gillson J L 1968 {\it Phys. Rev. Lett.} \textbf{21} 617
\bibitem{3} Adachi K, Sato K, Takeda M 1969 {\it J. Phys. Soc. Japan} \textbf{26} 631
\bibitem{4} Ogawa S, Waki S, Teranishi T 1974 {\it Int. J. Magnetism} \textbf{5} 349
\bibitem{5} Adachi K, Ohkohchi K 1980 {\it J. Phys. Soc. Japan} \textbf{49} 154
\bibitem{6} Hiraki H, Endoh Y, Yamada K  1997 {\it J. Phys. Soc. Japan} \textbf{66} 818
\bibitem{7} Yomo S 1979 {\it J. Phys. Soc. Japan} \textbf{47} 1486
\bibitem{8} Hiraka H, Endoh Y 1996 {\it J. Phys. Soc. Japan} \textbf{65} 3740
\bibitem{9} Sato T J, Lynn J W, Hor Y S, Cheong S W 2003 {\it Phys. Rev. B} \textbf{68} 214411
\bibitem{10} Barakat S, Braithwaite D, Alireza P, Grube K, Uhlarz M, Wilson J, Pfleiderer C, Flouquet J, Lonzarich G 2005 {\it Physica B} \textbf{359-361} 1216
\bibitem{11} Otero-Leal M, Rivadulla F, Garc\'{\i}a-Hern\'{a}ndez M, Pi\~{n}eiro A, Pardo V, Baldomir D, Rivas J 2008 {\it Phys. Rev. B} \textbf{78} 180415(R)
\bibitem{12} Sidorov V A, Krasnorusski V N, Petrova A E, Utyuzh A N, Yuhasz W M, Lograsso T A, Thompson J D, Stishov S M 2011 {\it Phys.Rev. B} \textbf{83} 060412 (R)
\bibitem{Larson} Larson A C, Von Dreele R B 1987 General Structure Analysis System (GSAS), Los Alamos National Laboratory, 2000, Report LAUR 86-748
\bibitem{Toby} Toby B H 2001 {\it J. Appl. Crystallogr.} \textbf{34} 210
\bibitem{14Sidorov} Sidorov V A, Thompson J D, Fisk Z 2010 {\it J. Phys.: Condens. Matter} \textbf{22} 406002
\bibitem{15Petrova} Petrova A E, Sidorov V A, Stishov S M 2005 {\it Physica B} \textbf{359-361} 1463
\bibitem{16Eiling} Eiling A, Schilling J S 1981 {\it J. Phys. F: Met. Phys.} \textbf{11} 623
\bibitem{17} Machida A, Synchrotron Radiation Research Center, Japan  (unpublished)
\bibitem{13Ogawa} Ogawa S 1976 {\it J.Phys. Soc. Japan} \textbf{41} 462
\bibitem{18} Mohn P,{\it Magnetism in the Solid State, An Introduction} (Springer)
\bibitem{19} Moria T, Usami K 1980 {\it Solid State Comm.} \textbf{34} 95
\bibitem{20} Shiga M 1981 {\it J.Phys. Soc. Japan} \textbf{50} 2573
\bibitem{21} Mushnikov N V, Goto T, Andreev A V, Zadvorkin S M 2000 {\it Phil. Mag. B} \textbf{80} 81
\bibitem{22} Liu X B, Altounian Z 2007 {\it J. Appl. Phys.} \textbf{101} 09G511
\bibitem{23} Petrova A E, Stishov S M 2009 {\it J. Phys.: Condens. Matter} \textbf{21} 196001
\bibitem{24} Maddox B R, Yoo C S, Kasinathan D, Pickett W E, Scalettar R T 2006 {\it Phys.Rev. B} \textbf{73} 144111
\bibitem{25} Landau L D, Lifshitz E M  1980 {\it Statistical Physics, Part 1, 3rd ed.} (Pergamon, Oxford)
\end{thebibliography}
\end{document}